# Development and evaluation of copper-containing mesoporous bioactive glasses for bone defects therapy

J Jiménez-Holguín[a], S Sánchez-Salcedo*[a,b], M Vallet-Regí[a,b], A J Salinas[a,b]*

[a]*Dpt. Química en Ciencias Farmacéuticas, Facultad de Farmacia, Universidad Complutense de Madrid, UCM; Instituto de Investigación Hospital 12 de Octubre, imas12, 28040 Madrid, Spain.*
[b]*Networking Research Center on Bioengineering, Biomaterials and Nanomedicine, CIBER-BBN, 28040 Madrid, Spain.*

**Abstract**

Mesoporous bioactive glasses, MBG, are gaining increasing interest in the design of new biomaterials for bone defects treatment. An important research trend to enhance their biological behavior is the inclusion of moderate amounts of oxides with therapeutical action such as CuO. In this paper, MBG with composition (85-x)$SiO_2$–10CaO–5$P_2O_5$–xCuO (x = 0, 2.5% or 5 mol-%) were synthesized, investigating the influence of the CuO content and some synthesis parameters in their properties. Two batch were developed; first one using HCl as catalyst and chlorides as CaO and CuO precursors, second one, using $HNO_3$ and nitrates. MBG of chlorides batch exhibited calcium/copper phosphate nanoparticles, between 10 and 20 nm. Nevertheless, CuO-containing MBG of nitrates batch showed nuclei of metallic copper nanoparticles larger than 50 nm and quicker in vitro bioactive responses. Thus, they were coated by an apatite-like layer after 24 h soaked in simulated body fluid, a remarkably short period for MBG containing up to 5 % of CuO. A model, focused in the copper location in the glass network, was proposed to relate nanostructure and in vitro behaviour. Moreover, after 24 h in MEM or THB culture media, all the MBG released therapeutic amounts of $Ca^{2+}$ and $Cu^{2+}$ ions. Because the quick bioactive response in SBF, the capacity to host biomolecules in their pores and to release therapeutic concentrations of $Ca^{2+}$ and $Cu^{2+}$ ions, MBG of the nitrate batch are considered as excellent biomaterials for bone regeneration.

*Dr. Antonio J. Salinas
*E-mail addresss:* salinas@ucm.es

# 1. Introduction

Mesoporous bioactive glasses (MBG) in the $SiO_2$–$CaO$–$P_2O_5$ system are being widely investigated for bone diseases therapy, due to their amazing properties including ordered mesopores arrangements in a very narrow pore size distribution as well as enormous surface area and pore volume [1–3]. MBG belong to the family of bioactive glasses that have attracted much attention because of their capability to bond with bone and to stimulate osteogenesis and angiogenesis [4–7]. The reactivity of bioactive glasses in the biological environment begins with the ionic interchange of ions in the glass and protons from medium. A silica-rich layer is formed in the glass surface that attracts $Ca^{2+}$ and $P(V)$ ions from medium producing an amorphous calcium phosphate layer that, after crystallizing into bone-like carbonate apatite, plays an important role in the bond of bioactive glasses to bone [8]. These physical-chemical processes are speeded by the huge textural properties of MBG. Moreover, these glasses are able to host molecules with biological activity inside their mesopores.

In the last years, several strategies were proposed to improve MBG by loading with small biomolecules and drugs and/or oxides of chemical elements with biological activities such as copper [9–12]. It must be considered that copper is an essential trace element that account between 1.5 and 2.1 mg per kg of human body [13]. Many studies demonstrated that copper lack produces a loss of bone mineral density [14,15]. This was explained by the loss of activity of Cu-containing enzymes, like superoxide dismutase with antioxidant properties or lysyl oxidase, involved in the collagen cross-linking during the bone formation [16,17]. Furthermore, it is known the role of copper mimicking a hypoxia situation by stabilizing hypoxia inducible factor-1α (HIF-1α) expression. This situation increases the transcription of hypoxia sensitive genes, like vascular endothelial growth factor (VEGF), the proliferation of endothelial cells and the recruitment and differentiation of mesenchymal cells [18–22]. In addition, copper shows antibacterial behaviour against *E. coli*, *B. subtilis*, *S. aureus* [23–25], and antiviral activity against norovirus [26] and influenza A virus [27].

A recent paper Romero-Sanchez et al. reported the in vitro and in vivo angiogenic capability of $SiO_2$–$CaO$–$P_2O_5$ MBG before and after doping with 5 % of CuO [28]. On the other hand, in a previous article of our research group, the synthesis conditions were adjusted to obtain MBG composites showing nanometric calcium phosphate nuclei in the glass walls [29]. These nanocomposites maintained the structural properties of MBG with enhanced in vitro reactivity in simulated body fluid (SBF) due to the presence of mentioned apatite-like nuclei. The presence of the new-formed nanoapatite layer was detected after 24 h of immersion.

The starting point of the present study was to obtain MBG with enhanced biological properties by the inclusion in their composition of 2.5 or 5 mol-% of CuO. Nevertheless, our previous studies demonstrated that the inclusion of a fourth oxide in ternary $SiO_2$–CaO–$P_2O_5$ MBG produced a decrease in the in vitro bioactive response in SBF [30]. For this reason, a revision and reinvention of the method of synthesis used by Cicuendez *et al.* for ternary MBG [29] was here proposed for the synthesis of quaternary $SiO_2$–CaO–$P_2O_5$–CuO MBG. Cicuendez *et al.* used hydrochloric acid and the chloride precursors of CaO and CuO for the synthesis of the nanocomposites because of the easy removal of chloride anions by calcination. Accordingly, we have investigated a first batch of CuO-containing MBG synthesised in the presence of chlorides. However, most research groups use more often nitrates as sources for the MBG syntheses [31–34]. Thus, a second batch of CuO-containing MBG was investigated employing nitric acid as catalyst and nitrates as CaO and CuO sources.

Therefore, two batch of MBG with 0, 2.5% or 5 mol-% of CuO will be developed in this study. First batch will use $HNO_3$ (Cu_HNO$_3$) as catalyst and nitrates as sources of CaO and CuO for the syntheses; the second batch HCl (Cu_HCl) and chlorides. A detailed study of MBG nanostructure materials, in vitro behavior in SBF and in common culture media of cells (MEM) and bacteria (THB) will be carried out. Our final goal will be to relate the in vitro behavior of the MBGs doped or not with copper to their nanostructure and determine if any of the MBG investigated could be of interest as a bone substitute.

## 2. Experimental

*2.1 Synthesis of the mesoporous bioactive glasses*

Two batch of MBG with composition (85-x) $SiO_2$–CaO–$P_2O_5$–xCuO, (x= 0, 2.5 and 5) were synthesized by the EISA (Evaporation-Induced Self Assembly) method. First batch, using $HNO_3$ 1M as catalyst and $Ca(NO_3)_2·4H_2O$ and $Cu(NO_3)_2·2.5H_2O$ as sources of CaO and CuO. Second one, using HCl 1 M, $CaCl_2·2H_2O$ and $CuCl_2$ as catalyst and sources of CaO and CuO. In all cases $SiO_2$ and $P_2O_5$ sources were tetraethyl orthosilicate (TEOS) and triethyl phosphate (TEP), respectively. In brief, the synthesis was performed by mixing Pluronic® F127, ethanol (99.98%), distilled water and HNO$_3$ or HCl, in a flask that was covered with Parafilm® and kept 1 h under stirring. (Reactants were purchased from Sigma-Aldrich, St. Louis, MO, USA). The corresponding amounts of reactants were added in the sequence indicated in the **Table 1** with 60 min intervals. The mixture was kept 14 h at 40 °C under stirring. Then, the mixture was poured in open Petri dishes (27 mL/plate) and kept 4 d at 30 °C in a stove to evaporate the ethanol formed in the hydrolysis of the alkoxides. Along this process the critical micellar concentration that allowed obtaining the mesoporous phase was reached, and the extra drying produced dried gels with sheets shape with the desired composition. MBG powders were obtained after heating the sheets for 6 h at 700ºC (heating ramp of 1 °C/min), grinding and sieving with a 40 μm mesh. The in vitro studies were performed in disks, 10 mm diameter, 1 mm height, obtained compacting 100 mg of MBG powders with 5 MPa of uniaxial pressure for 1 min.

*2.2. ysicochemical Characterization of Samples*

Samples were characterized by Small-Angle X-ray diffraction, SA-XRD, in a X'pert-MPD system (Eindhoven, The Netherlands) equipped with Cu Kα radiation in the 0.6 to 8º 2θ range; Thermogravimetric and Differential Thermal analysis (TG/DTA) in the 30 °C to 900 °C interval (air flow: 100 mL/min) in a Perkin Elmer Pyris Diamond system (Waltham, MA, USA), Fourier transformed infrared (FTIR) spectroscopy in a Thermo Scientific Nicolet iS10 apparatus (Waltham, MA, USA) equipped with a SMART Golden Gate attenuated total reflection ATR diffuse reflectance accessory and Transmission Electron Microscopy (TEM), in a JEM-2100 JEOL microscope operating at 200 kV (Tokyo, Japan). For TEM analysis, samples were dispersed in ethanol, sonicated and deposited in a nickel grid coated with a holed polyvinyl-formaldehyde layer. Moreover, materials were characterized by nitrogen porosimetry using Micromeritics 3 Flex (Norcross, GA, USA) for these measurements, samples were previously degassed 24 h at 120 °C under vacuum. Surface area, $S_{BET}$, was calculated by the Brunauer-Emmett-Teller (BET) method [35], and pore size distributions by the Barret–Joyner–Halenda (BJH) method [36]. Finally, MBG samples were analyzed by $^{29}$Si and $^{31}$P solid state single pulse magic angle spinning nuclear magnetic resonance (SP MAS-NMR) on a Bruker Avance AV-400WB spectrometer (Karlsruhe, Germany) equipped with a 4 mm zirconia rotor. Frequencies were set at 79.49 and 161.97 MHz for

$^{29}$Si and $^{31}$P respectively and chemical shift values were referenced to tetramethylsilane, for $^{29}$Si, and $H_3PO_4$, for $^{31}$P. Time period between accumulations were 5 s, for $^{29}$Si and 4 s, for $^{31}$P, and the number of scans was 10,000.

*2.3. In vitro studies in acellular solutions*

The reactivity of MBG was assessed under in vitro conditions in simulated body fluid to evaluate the in vitro bioactivity and in culture media of cells or bacteria to quantify the amount of inorganic ions released to medium.

*2.3.1. Assays in SBF*

The surface reactivity of the samples was assessed by following the formation of an apatite-like layer on the MBG disks surface after being immersed in SBF [37]. MBG disks were previously sterilized for 20 min under UV radiation (10 min/face) in a laminar flux cabinet and the SBF solution was filtered through a 0.22 μm filter to avoid bacterial contaminations. MBG disks were soaked for 8 h, 24 h, 3 d and 7 d in 30 mL of SBF at 37ºC, pH 7.4. The volume of SBF was chosen according our previous studies where we always maintain constant the proportion $V_S = D_S/ 0,075$, being $V_s$ the SBF volume in mL and $D_S$ the geometric external area in cm$^2$ of the item investigated [31]. Inside the container disks were maintained in a vertical position with a platinum wire. Assays were performed with two replicas by material and a control of material-free SBF.

Apatite-like phase formation was assessed by wide angle XRD (WA-XRD, 2θ from 10-70 °), FTIR spectroscopy and scanning electron microscopy and (SEM) in a JSM-7600F (JEOL) microscope (Tokyo, Japan) coupled with an EDS spectroscopy system (Oxford Instruments, Abingdom, UK).

*2.3.2 Ions release from disks*

The assay was performed by soaking MBG disks in 2 ml of Eagle's minimal essential medium (MEM) with 0,5% of penicillin-streptomycin, and, independently, in 2 ml of Todd Hewitt Broth (THB) medium (both from Sigma-Aldrich. St. Louis, MO, USA), to explore the ion release in mammal cell culture medium and bacterial culture medium. Samples were collected at 8 h and 1, 2, 3, 4, 5 and 6 d. The accumulative concentration of $Ca^{2+}$, P(V) and $Cu^{2+}$ in both medium was determined by inductively coupled plasma/ optical spectrometry (ICP/OES) using an OPTIMA 3300 DV device

(Perkin Elmer, Waltham, MA, USA). Two replicas were used of each composition. Three measures were taken of each replica to do statistical analysis.

*2.3.3 Statistical Analysis*

Results were evaluated with nonparametric Kruskal-Wallis test and post-hoc Dunn´s test. A probability value, p-value < 0.05 was considered significant.

## 3. Results and Discussion

*3.1 Glass powders characterization*

MBG powders obtained were characterized by TEM and SA-XRD to determine their nanostructure and check their mesoporous structure. EDS and FTIR spectra were carried out to determine the composition and to confirm the complete removal of the surfactant and other possible impurities during the calcination step. Finally, $^{29}$Si and $^{31}$P NMR spectroscopy determine the glass network forming species at atomic level which will allow understanding the glasses reactivity in aqueous solution.

**Fig. 1** depicts the TEM images and EDS spectra of the MBGs investigated. In all cases the presence of ordered mesoporosity is shown. It is remarkable that whereas MBG of the Cu_HNO$_3$ batch show highly ordered arrangements of mesopores channels, the ones of Cu_HCl batch exhibit somewhat curved channels. Furthermore, both batch of materials showed the presence of pseudo-spherical nanospheres fairly different for each batch. Thus, in Cu_HCl batch, were detected abundant nanoparticles under 20 nm with calcium, phosphorous and oxygen composition, similar to apatite nanoparticles previously reported by our group [29]. However, in Cu_HNO$_3$ batch Cu enriched MBGs, smaller amount of bigger size nanoparticles was detected. More than 30 nanoparticles were measured. The average size of the nanoparticles was 59±13 and 77±36 nm, respectively for MBG with 2.5% and 5% of CuO. **Table 2** collects the EDS analysis of the MBG glasses synthesized. A good agreement was observed between the experimental values and the nominal ones from the amounts of reactants used in the synthesis. ~~NMR spectroscopy results, which will be presented later on, will give some extra information about the nanoparticles formed in both batch of MBG.~~

**Figs. 2. A and A'** includes SA-XRD diffraction patterns of MBGs. As observed, both batch of materials exhibited a sharp maximum at around 1.0 ° in 2Θ, typical of mesoporous order, that could be assigned to the (10) reflection of a 2D p6mm

hexagonal phase (**Table 3**). In addition, in Cu_HCl MBGs it can be observed a shoulder at around 1.5 ° in 2Θ that was assigned to the (10), (20) reflections [30]. Nevertheless, in Cu_HNO$_3$ batch (10) and (20) reflections were less defined, suggesting a lower order of mesoporous for these samples. SA-XRD results agree with those observed by TEM showing that Cu_HCl samples exhibited higher mesoporous order than Cu_HNO$_3$ ones.

To determine the effect of catalyst and precursors based on it in the textural properties, MBG were analysed by nitrogen adsorption. **Fig. 2.B and B'** showed the isotherms of the MBG investigated. As observed, type IV isotherms, characteristic of mesoporous materials, were obtained in all cases. The hysteresis cycles were slightly different depending on the MBG composition. Thus, 2.5% and 5% Cu_HNO$_3$ MBGs present type H1 cycle of hysteresis, characteristic of materials with cylindrical pore channels or with agglomerates of spheres. In contrast, nitrate CuO-free glasses and all the Cu_HCl glasses showed cycle of hysteresis type H2 related with materials exhibiting pores with bottleneck shape [38]. **Fig. 2. C and C'** shown pore diameter distribution of MBG materials, centred in 6-6.5 for Cu_HNO$_3$ MBGs and in 8.5-9 for Cu_HCl MBGs.

**Table 2** also includes the textural properties of MBG powders and after be conformed into disks. As is observed, in CuO-free samples (0%Cu_HCl and 0%Cu_HNO$_3$), S$_{BET}$ were somewhat great for nitrate batch, but in CuO-containing MBG the chloride samples showed higher S$_{BET}$. In addition, Cu_HNO$_3$ glasses S$_{BET}$ decreased from 301 to 261 m$^2$/g and pore volume from 0.41 to 0.35 cm$^3$/g when CuO increased from 0 % to 5 %. These results agree with previously reported for the addition of copper to other MBG compositions [28,39–41]. In addition, pore diameter of Cu_HNO$_3$ samples slightly decreased from 6.7 to 6.1 nm with the CuO additions due to the segregation of metallic copper nanoparticles which are larger hinder the structural order of the MBG. However, in Cu_HCl samples when CuO increased from 0 % to 5 %, S$_{BET}$ increased from 218 to 295 m$^2$/g and pore volume from 0.40 to 0.47 cm$^3$/g. This effect can be explained due to higher formation of calcium/copper phosphate clusters decreases the proportion of network modifiers (Ca$^{2+}$ and Cu$^{2+}$) in the MBG structure increasing the participation of the network former (SiO$_2$) and therefore increases the order in the structure. Nevertheless, all Cu_HCl samples pore diameters were around 8 nm, showing that CuO inclusions have not effects on the pore diameter. In summary, nitrogen adsorption results suggested that copper exerts a different effect in the textural properties of MBG depending on the catalyst and precursors used for the synthesis.

$^{29}$Si NMR measurements were carried out to investigate the environments at the atomic level of the network former on the MBG materials (**Fig. 3**). In this manner $Q^2$, $Q^3$, and $Q^4$ represent the three different environments of silicon atoms (Si*): (NBO)$_2$Si*–(OSi)$_2$, (NBO)Si*–(OSi)$_3$, and Si*(OSi)$_4$ (NBO: non-bonding oxygen). **Table 4** includes the chemical shifts, CS, and the peak areas expressed in percentages. Silica network connectivity <$Q^n$> of each MBG was also included. As is observed, the signals at -111 to -112 ppm were assigned to $Q^4$, at -102 to -106 ppm to $Q^3$ and at -95 to -98 ppm to $Q^2$ [42].

**Fig. 4 and Table 4** shows the $^{31}$P NMR spectra of MBG, where $Q^0$ and $Q^1$ represent the phosphorus atoms (P*) in the PO$_4$$^{3-}$ species: P*–(NBO)$_4$ and (NBO)$_3$–P*–(OP). The intense signals at 1.8 to 2.2 ppm were assigned to the $Q^0$ environment typical of amorphous orthophosphate. For Cu_HNO$_3$ samples a second weak signal ($Q^1$), between -5.4 and -4.5 ppm, was observed which intensity increased as CuO in the MBG increased. This second maximum in the $Q^1$ tetrahedra range could be assigned to P–O–Si environments as previously reported [42]. This effect indicates a higher incorporation of phosphorous in the glass network and consequently a decrement of amorphous orthophosphate clusters in support of metallic copper nanoparticles formation as it is observed by TEM (**Fig. 1**). The very slight variation of network connectivity when copper is added to MBGs demonstrates the segregation of Cu in metallic nanoparticles and no copper participation in the MBG structure.

In addition, as observed in **Table 4,** high values of $Q^0$ for Cu_HCl glasses, up to 91.7 % and 90.4 %, indicate that phosphorous would be mainly in these glasses as orthophosphate clusters decreasing the proportion of P–O–Si units. Besides, it is worthy to note that the introduction of copper in Cu_HCl samples caused a large decrease in the percentage of $Q^1$, from 29.9%, for CuO-containing glass, to 8.3 % and 9.1 % respectively, for MBG containing 2.5 and 5% of CuO.. This is in agreement with the decrees of network connectivity when Cu is incorporated [43], indicating this ion act as a network modifier and favouring the orthophosphate clusters that were observed by TEM (**Fig. 1**). Moreover, these values suggest that copper hinders phosphate groups to participate as network formers which will be affect the in vitro bioactive response of Cu_HCl MBGs.

From these results we can conclude that the differences observed in the $Q^0$-$Q^1$ balance of the MBGs with 0% of CuO, may be due to the fact that, firstly, the phosphorous was integrated with a smaller amount in HNO$_3$ batch than in the HCl batch as shown in the

**Table 2**, and as reported by Cicuéndez *et al.*, these materials showed nanoparticles of calcium phosphate in their structure after their synthesis. On the other hand, copper is acting in different ways in each batch, favouring the participation of phosphorous as a network former in $HNO_3$ batch with higher percentage of $Q^1$, while in the HCl batch it seems to participate as ortophosphate clusters with higher percentage of $Q^0$.

Previous results have shown a detailed nanostructural characterization of the mesoporous materials synthesised. Materials obtained as powders were shaped as disks to investigate their bioactive response in a simulated body fluid and their capability to release ions with therapeutic activity to the surrounding medium. We will try to relate the results of the in vitro studies with the MBG nanostructure previously described.

*3.2 Characterization of the MBG disks*

In vitro bioactivity and ion release assays were carried out using disks of MBG powders as described in the experimental section. Previously, disks were characterized by nitrogen adsorption to check that the processing of powders did not modify significantly the textural properties of MBG. **Fig. 5** shows the isotherms and the pore diameter distributions of MBG disks. As is observed, these curves are analogous to those of corresponding MBG powders that were shown in **Fig. 2.** Moreover, **Table 2** collects the textural parameters of both MBG powders and MBG disks, obtained from the corresponding isotherms. Values in **Table 2** confirms that both batchs of MBG powders lose rather little surface area in their processing into disks, 17 % in average for $Cu\_HNO_3$ glasses, and 3 % for Cu_HCl samples. These decreases are consequence of the partial collapse of the pore structure of glasses. However, the isotherms shape and the pore diameter size remain stable for both batch of materials. Furthermore, the $S_{BET}$ of MBG disks and pore volume are in all cases high enough for the intended application of these materials for bone disease treatment.

*3.3 In vitro bioactivity assays of the MBG disks in simulated body fluid*

Two disks of each MBG composition were soaked in SBF for 8 h, 24 h, 3 d and 7 d. Disks were analysed by FTIR (**Figs. 6 and 7**) and SEM-EDS (**Fig. 8**) before and after soaking. FTIR spectra of MBG before soaking in SBF showed asymmetric bending vibration Si–O–Si bands at 1100-1040 cm$^{-1}$ and 478-441 cm$^{-1}$ and another band at 800-

793 cm$^{-1}$ of the symmetric Si–O stretch [31]. Moreover, in the spectra of Cu_HNO$_3$ glasses, a very low intensity band was present about 590 cm$^{-1}$ that can be attributed to the presence to some phosphate phase with extremely low crystalline size. However, the spectra of Cu_HCl samples exhibited in this area two small bands at around 600 and 570 cm$^{-1}$ characteristic of phosphate in a crystalline environment. Such bands, visible before soaking in SBF, were identified by Cicuéndez et al. as the nanocrystalline apatite phases present their nanocomposites [29].

After soaking nitrate-based MBG in SBF were detected bands at 600 and 570 cm$^{-1}$ which intensity increased with soaking time. These bands, well known by our research group, are indicative of an in vitro bioactive response of tested material due to the nanohydroxycarbonate apatite formation in SBF [30]. After 8 h of soaking these bands were clearly visible in 0%Cu_HNO$_3$ and 2.5% Cu_HNO$_3$ spectra, although the intensity decreased when the CuO content increased. Thus, for 5%Cu_HNO$_3$, the bands were only visible after 24 h in SBF. This delay in the nanohydroxycarbonate apatite formation was explained considering that the high amount of copper in the glass prevents a quicker bioactive response. The very short time for the in vitro bioactive response of the CuO-free glass is similar to reported by our group and other research groups in analogous ternary MBG. However, 8 h to show in vitro bioactive response can be considered an extremely short time for a MBG containing 2.5% of CuO.

On the other side, MBG of the Cu_HCl series showed a more moderate in vitro bioactive response than Cu_HNO$_3$ glasses. Although, all the glasses exhibited an in vitro bioactive response after less than 7 d in SBF. In some of the spectra of Cu_HCl MBGs after soaking in SBF is not clear about whether the two bands in this region come from the phosphate nuclei initially present in the MBG or of the apatite-like formed after in vitro bioactivity tests. Probably in several cases the bands observed are sum of the two types of bands mentioned. Therefore, another technique is necessary, i.e. SEM, which allowed us to check in which cases the MBG surface was coated by a newly formed material.

**Fig. 8** shows the SEM images of MBG disks of the two series of MBG investigated before and after be soaked 8 and 24 h in SBF. For each sample, the corresponding EDS spectrum was also included, as well as the Ca/P molar ratio. As is observed, the images of MGB disks before soaking show the bare surface ~~of the disks~~, where the grains of the MBG powders used to obtain the disks are still visible. However, after 8 h of soaking, the surface of more reactive samples appeared coated by a newly formed layer, whereas the surface of other samples remained unchanged. In this sense, two trends were observed: for the same composition, nitrate-based samples exhibited always a quicker bioactive response than chlorides MBG batch ones, and for both series, this response decreased when MBGs doped with copper increased. It should be noted that copper free MBG samples seem to show a thin layer formed by small needles shaped nanocrystalline at 8 h of submersion in SBF, indicating that this layer had to be produced quite a long time before 8 h. This result along with the composition of indicated by EDS is in agreement with two bands at around 600 and 570 cm$^{-1}$ characteristics of phosphate in a crystalline environment those observed by FTIR spectra (**Figure 6)** and can be unequivocally identified as the apatite-like layer. After 24 h of testing the surface of all of Cu_HNO$_3$ batch samples were covered with the same thick layer of nanocrystals observed at 8h for the CuO free samples indicative of a positive bioactive behaviour. Nevertheless, in the Cu_HCl batch this thick new formed layer was only observed at 24 h for the Cu-free sample. However, in the samples with 2.5% and 5% of CuO the layer formed was less thick and did not totally coat the surface of disks, which is a clear indication that these two samples are bioactive but a little less than the Cu_HNO$_3$ batch.

This observed mineralization process is well known in the literature [44–46], characterized by the early formation of ovoalated nuclei of amorphous calcium phosphate as observed for 2.5% and 5%_HNO$_3$ at 8 h and can not be observed for 0%_HNO$_3$ due to their quick in vitro response. This phase should have occurred between 8 and 24 h for the Cu_HCl batch because cannot be appreciate it at 8h. Subsequently, maturation of this layer to octacalcium phosphate should take place between 8 and 24 hours for the the Cu_HNO$_3$ batch and HCl_Cu-free, resulting in the formation of the calcium-deficient hydroxycarbonate apatite layer, as observed by the appearance of the needle shapes **(Figure 8).** This process seems to slow down but not hinder for copper-enriched samples for Cu_HCl batch.

Although MEM and THB are not the ideal solution to evaluate the in vitro bioactivity of a glass, which is generally performed in SBF, it has been decided to carry out these studies because that were the media used for the mammalian cell and bacteria culture studies and it was necessary to know the surface reactivity of the MBG samples investigated. **Figs. 9 and 10** show the cumulative Ca, Cu and P ions release from both MBG batchs in MEM and THB culture media up to 6 d. Taking into account that MEM present 90% more amount of Ca and 50% less amount of P in its initial composition than THB absolute values between them are not comparable. In case of Cu_HNO3 batch Ca and P ions release are typical of high bioactive glasses followed the release-precipitation equilibrium (**Fig. 9**). In MEM the three nitrate-based batch released analogous Ca amounts, being higher in CuO-free MBG sample after 3d for both media and significantly different in MEM, reaching 102 ppm for Cu-free sample (**Fig. 9**). Regarding Cu ions (**Fig. 9**), knowing that copper are imbibed in glass structure as metallic nanoparticles (**Figure 1 and 3**), higher amount of Cu nanoparticles in MBGs increase the release to the medium as it can be seeing in MEM, being the difference between 2.5% Cu_$HNO_3$ and 5% Cu_$HNO_3$ samples up to 300 ppm after 6 d of treatment. However, was inversely proportional due to the precipitation of copper phosphate when the cooper in medium is increased in THB (**Fig. 9**) [30].

The same assay was performed for Chloride-based batch (**Fig. 10**). Since Cu_HCl batch stands out for having a high content of Ca/CuP nanoparticles imbibed in their structure (**Fig. 1**) and copper ions as a network modifier as it have been demonstrated by NMR (**Fig. 3, Table 4**). Ca and P ions (**Fig. 10**) release are typical of bioactive glasses followed the release-precipitation equilibrium. Moreover, there was a significantly deference between CuO-free samples and MBGs doped with Cu due to the different bioactive kinetic that has been demonstrated in the bioactive behaviour of this Chloride-based batch (**Fig. 8**) being more bioactive the CuO-free MBGs. In case of Cu ion (**Fig. 10**) release when copper enrichment are higher in the Cu_HCl batch the amount of Cu released are significantly increased in both cell and bacteria media with a difference up to 150 and 600 ppm in MEM and THB, respectively. This effect confirms the decrease network connectivity ($Q^n$) calculated in NMR (**Table 4**) when Cu are increased in the MBG structure and thus more soluble in aqueous media.

It is remarkable that all CuO-containing MBG samples reached therapeutic concentration of Cu ions before 24 h of release in MEM medium and their increment of Cu releases by each time are cytocompatible [2]. Furthermore, all CuO-containing MBG samples released more amounts of $Cu^{2+}$ ions in THB than MEM after 6 d of assay, being an excellent result because of their therapeutically application against bacterial infection. However, small amount of MBG enriched with copper will be investigated to ensure their cytocompatibility along with their antibacterial properties.

**4. Conclusions**

Two batch of mesoporous bioactive glasses with composition (85-x)$SiO_2$–10CaO–5$P_2O_5$–xCuO (x= 0, 2.5%, 5 mol-%) have been obtained by using nitrate or chloride precursors in their MBG syntheses. Nanostructure and in vitro behaviour of MBG were dependent on the synthesis conditions. In summary, nitrogen adsorption results suggested that copper exerts a different effect in the textural properties of MBG depending on the catalyst and precursors used although a high pore volume and specific surface area have been maintained in MBGs with and without copper. Thus, transmission electron microscopy revealed in MBGs highly ordered mesoporous nanostructure with calcium/copper phosphate nanoparticles (10-20 nm) in Cu_HCl batch and nanoparticles larger than 50 nm in size identified as metallic Cu nuclei in Cu_$HNO_3$ batch.

The different nanostructures of glasses yielded to different in vitro behaviours in simulated body fluid (SBF) or in cell (MEM) and bacteria (THB) culture media. A model, focused in the copper location in the glass network, was proposed relating nanostructure and in vitro behaviour. At 24 h, all the MBG investigated were able to release therapeutic amounts of $Ca^{2+}$ and $Cu^{2+}$ ions. Moreover, Cu_$HNO_3$ batch exhibited and quicker in vitro bioactive responses. Thus, MBGs with 2.5% of CuO was coated by the apatite-like layer after 24 h in SBF, a remarkably short period for a CuO-containing MBG. Because of the quick bioactive response in SBF, the capacity to host biomolecules and to release therapeutic concentrations of ions, MBG of the nitrate batch are proposed as biomaterials for bone regeneration, and it is considered that they deserve further in vitro and in vivo studies.

## 5. Acknowledgements.

This study was supported by research grants from Instituto de Salud Carlos III (PI15/00978) project co-financed with the European Union FEDER funds, the European Research Council (ERC-2015-AdG), Advanced Grant Verdi-Proposal No.694160 and Youth Employment initiative (YEI) from Comunidad de Madrid (PEJ-2017-2017-AI/SALl-5825).

## 6. Declaration of Competing Interest.

I declare that (1) the article is original and unpublished and has not or is not being considered for publication elsewhere, (2) all authors have seen and approved the manuscript, and (3) the authors declare that they have no known competing financial interests or personal relationships that could have appeared to influence the work reported in this paper.

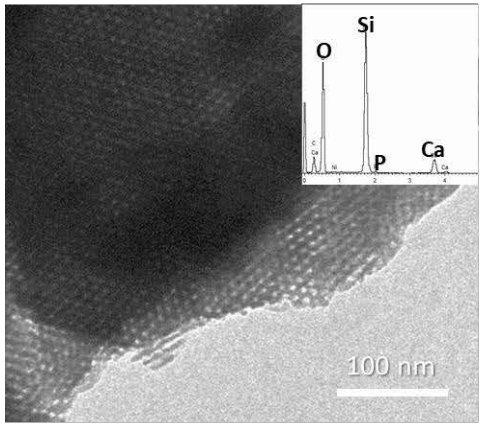 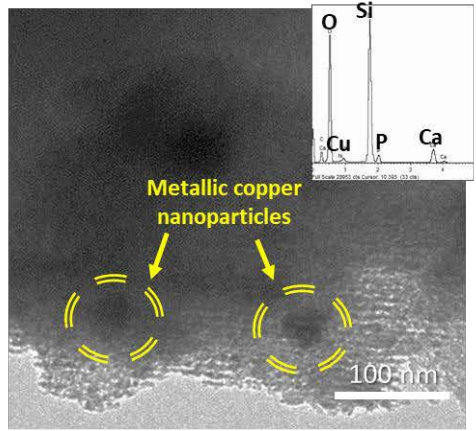 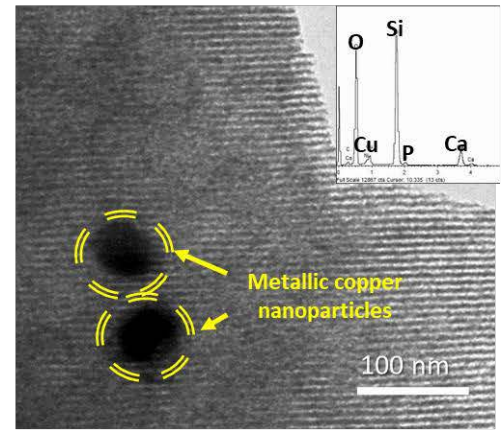
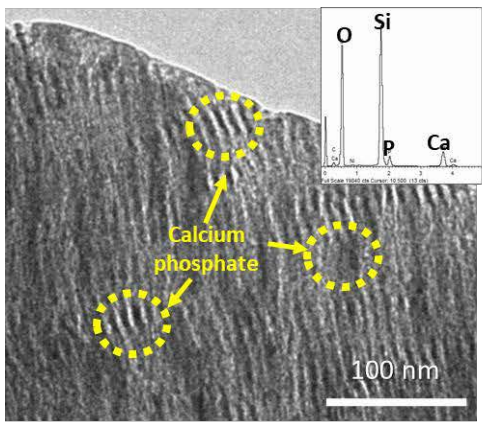 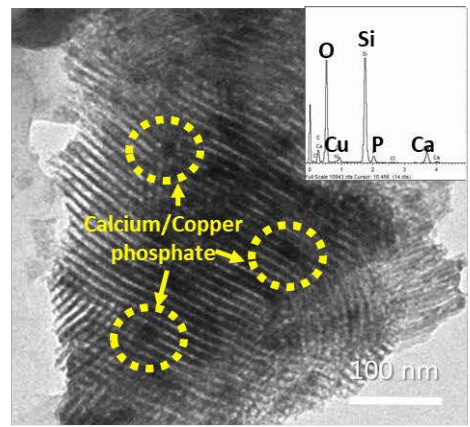 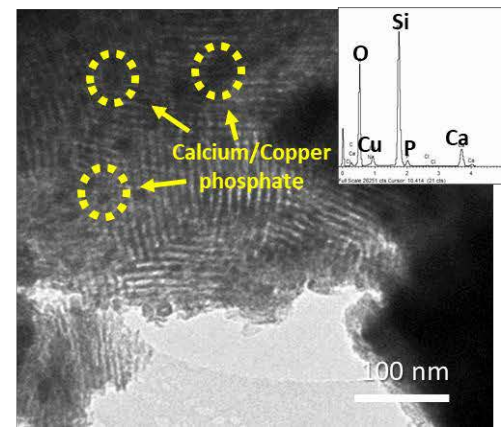

**Figure 1.** TEM images of samples enriched with 0, 2.5 and 5% of CuO synthetized using $HNO_3$ or HCl as catalyst. Highlighted with striped circles the metallic copper nanoparticles and with dotted circles those of calcium / copper phosphate. Each image included in the up right corner their respective EDS espectrum.

(2-column)

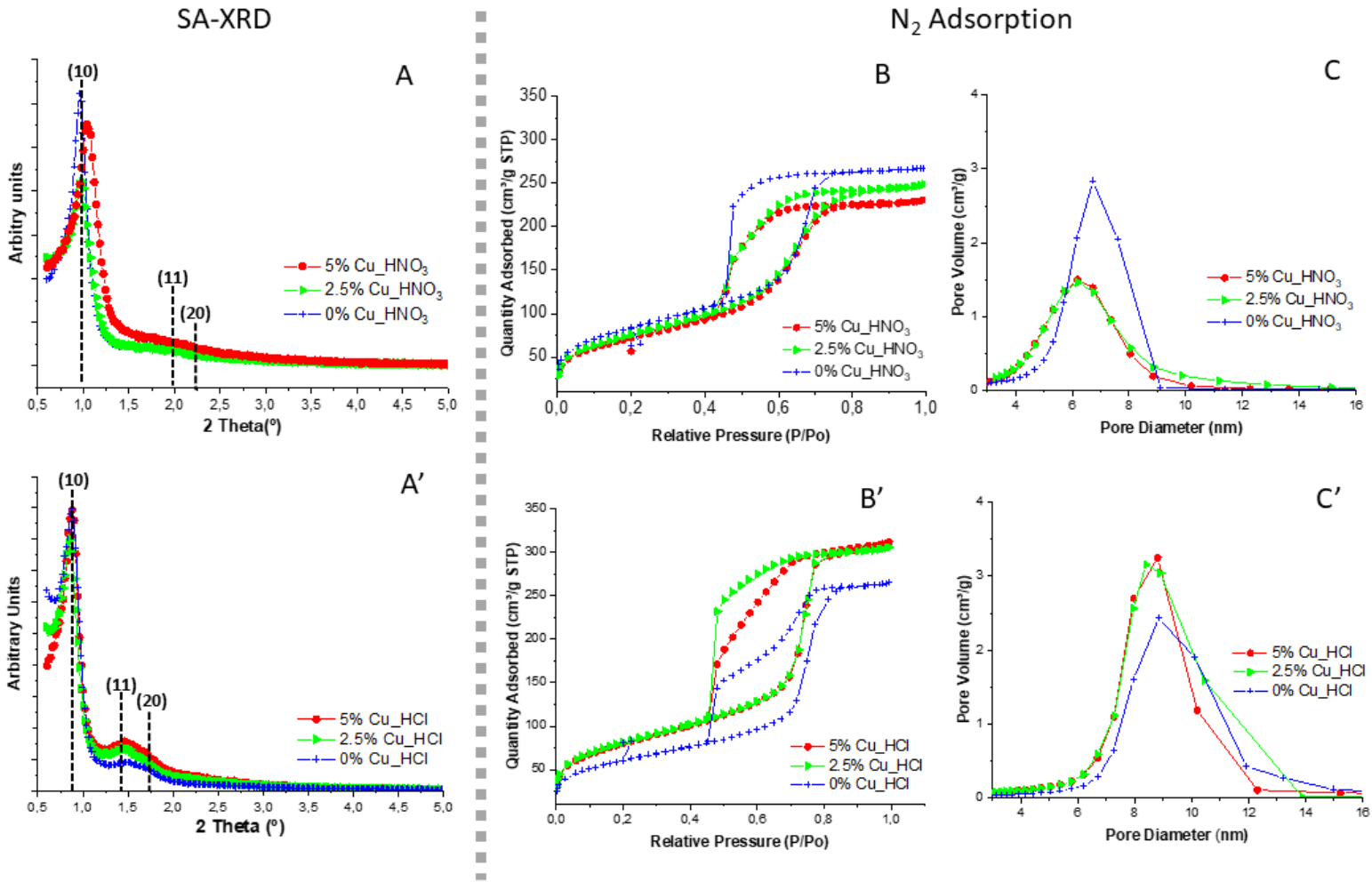

**Figure 2.** Left, SA-XRD patterns of MBG synthetized in the presence of nitrates or chlorides (A, A'). Right, $N_2$-adsorption results showing the isotherms (B, B') and pore size distributions (C, C') of both MBG batchs.

(2-column)

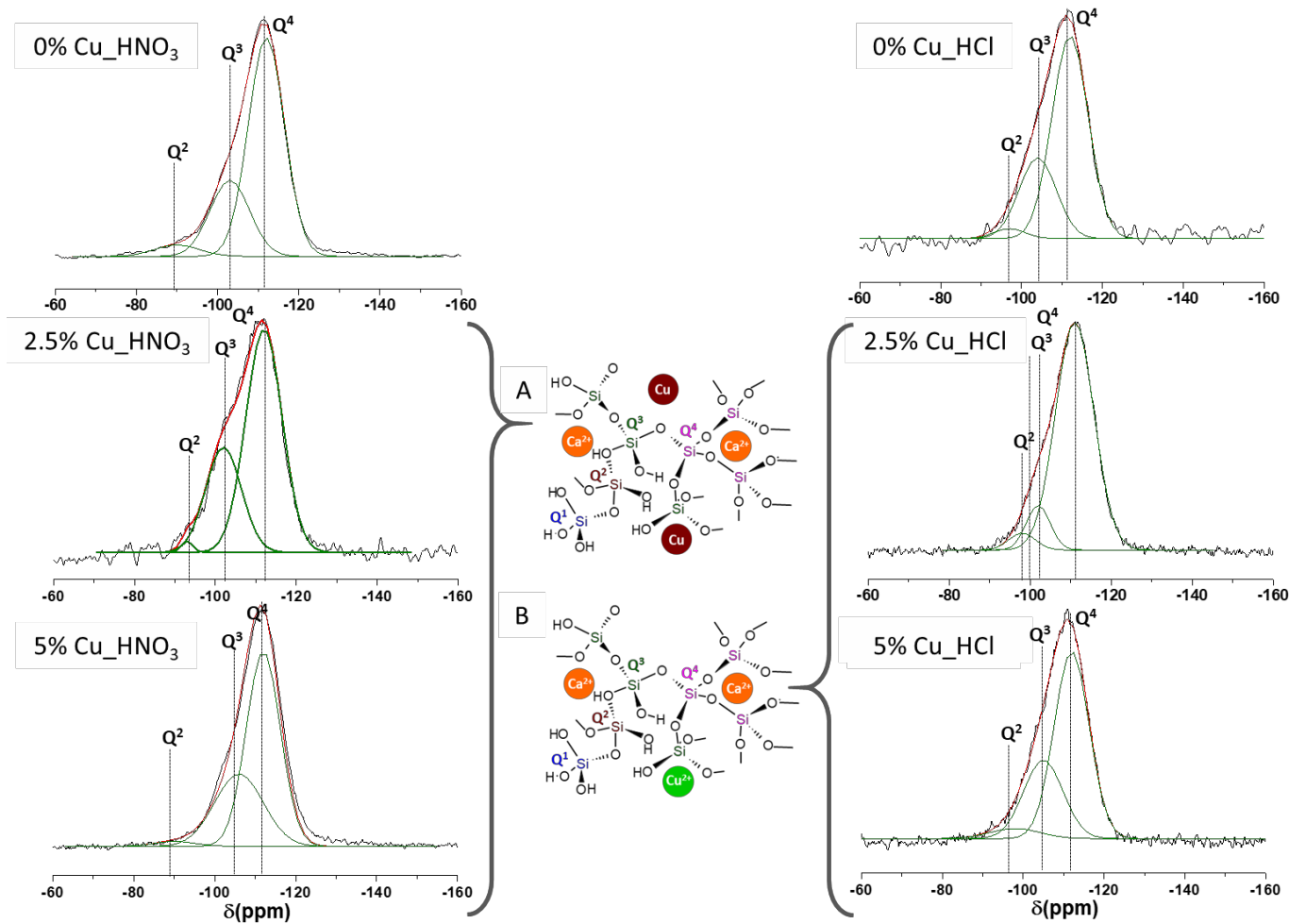

**Figure 3.** Solid-state $^{29}$Si single-pulse MAS-NMR spectra of MBG. $Q^2$, $Q^3$ and $Q^4$, obtained by Gaussian line-shape deconvolution, are displayed in green. Left: nitrate batch. Right: chlorides batch. Two models showing the copper location in the glass network were proposed. (A) Nitrate-batch. (B) Chloride-batch. The error associated with the spectra fitting is indicated as a red line.
(1.5-column)

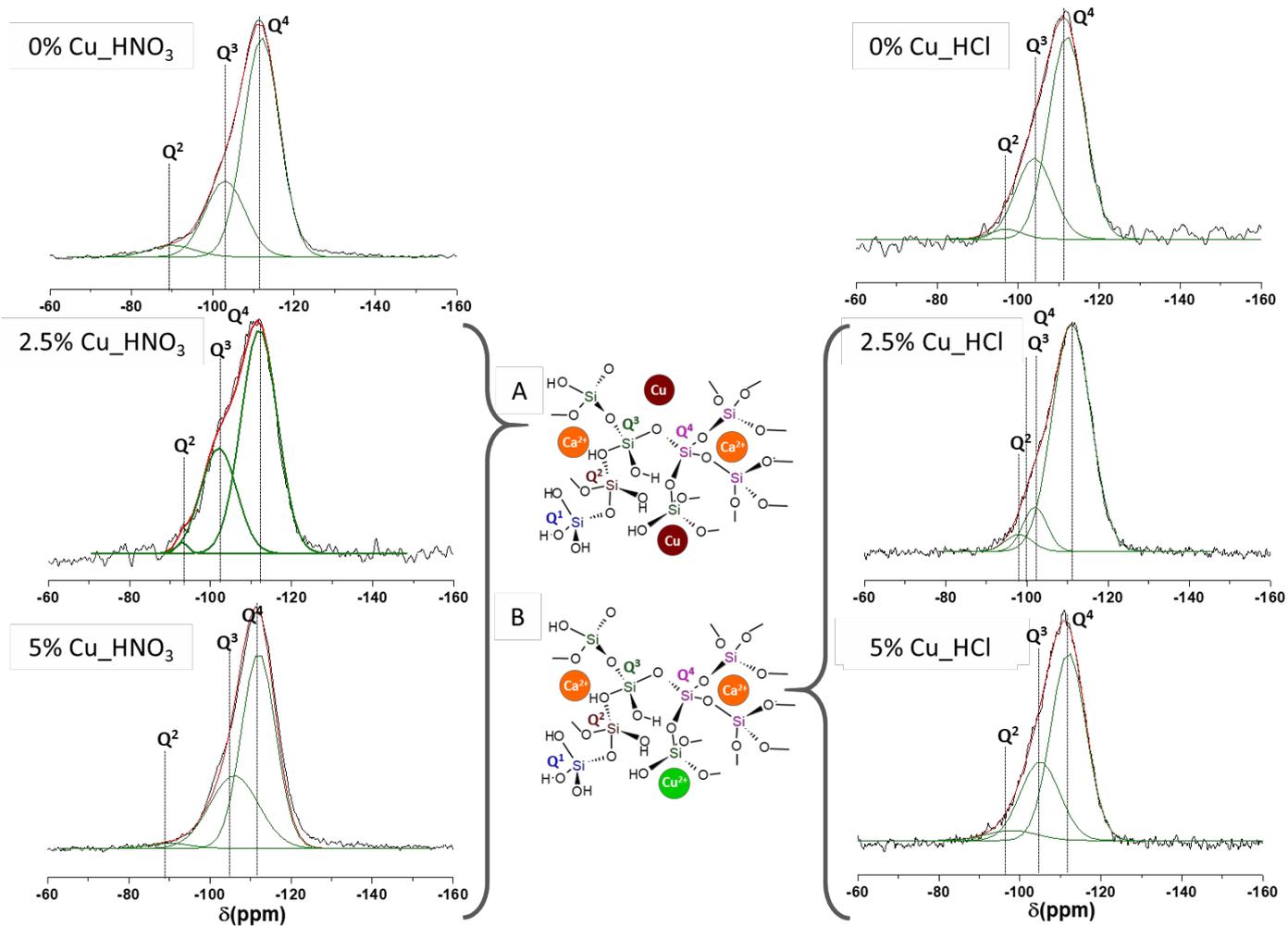

**Figure 4.** Solid-state $^{31}$P single-pulse MAS-NMR spectra of MBG $Q^0$, and $Q^1$, obtained by Gaussian line-shape deconvolution, are displayed in green. Left: Nitrate batch. Right: Chlorides batch. The error associated with the spectra fitting is indicated as a red line.
(1.5-column)

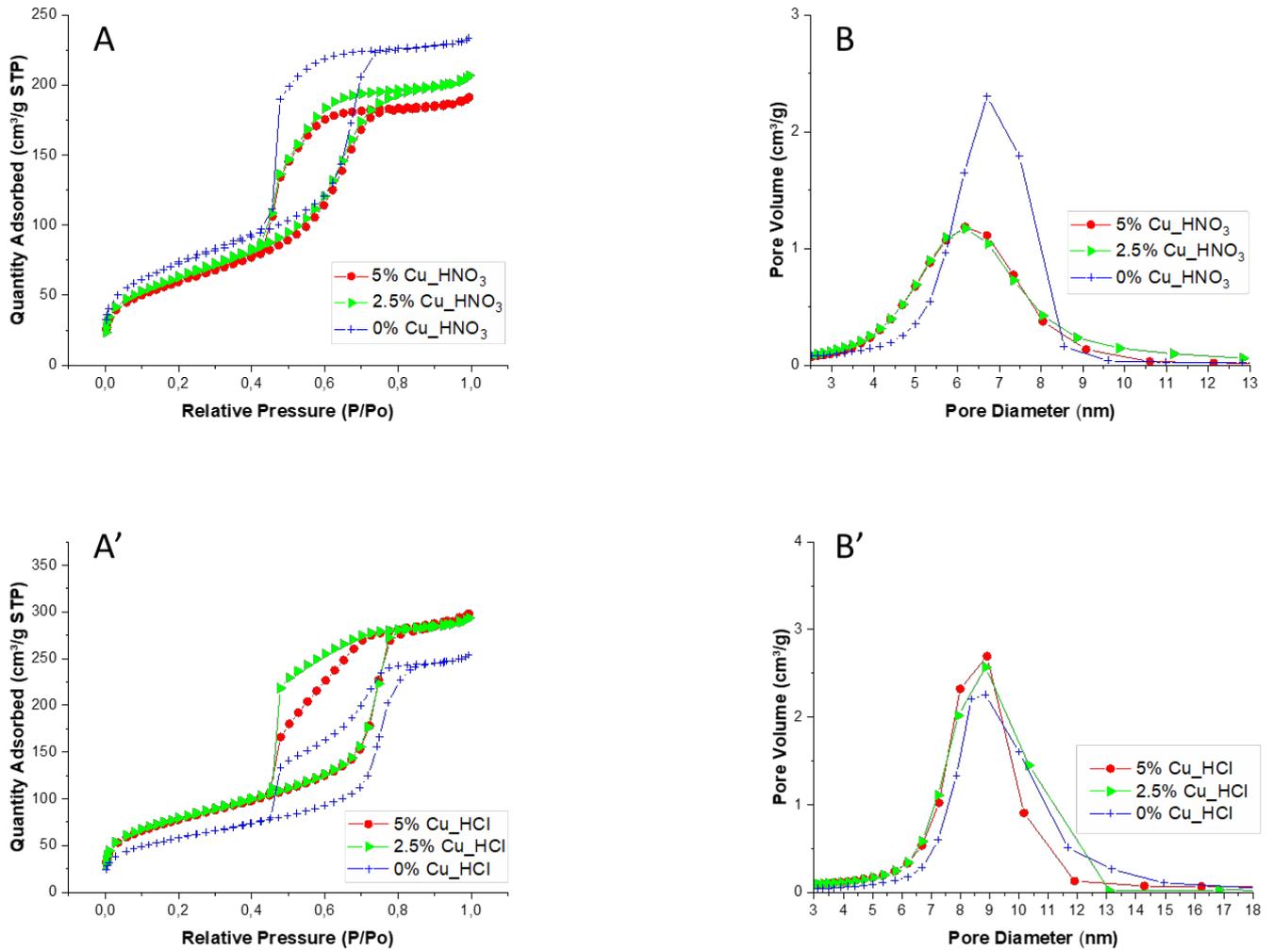

**Figure 5.** N$_2$-adsorption results showing the isotherms (A, A') and pore size distributions (B, B') of both MBG batch after be compacted into disks.

(1.5-column)

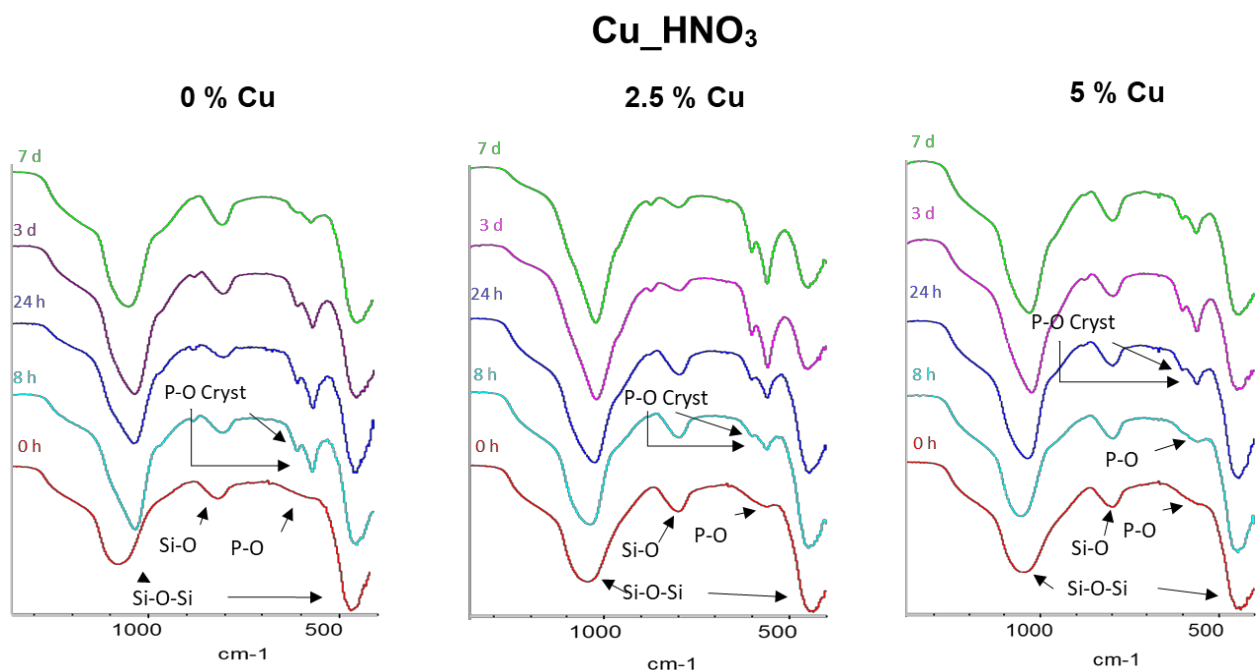

**Figure 6.** FTIR spectra of nitrate batch before and after 8 h, 24 h, 3 d and 7 d of immersion in simulated body fluid. Arrows highlight the most significant bands of Si-O and P-O in the spectra. Of interest are the bands of phosphate in a crystalline environment, labeled as P-O Cryst.

(1.5-column)

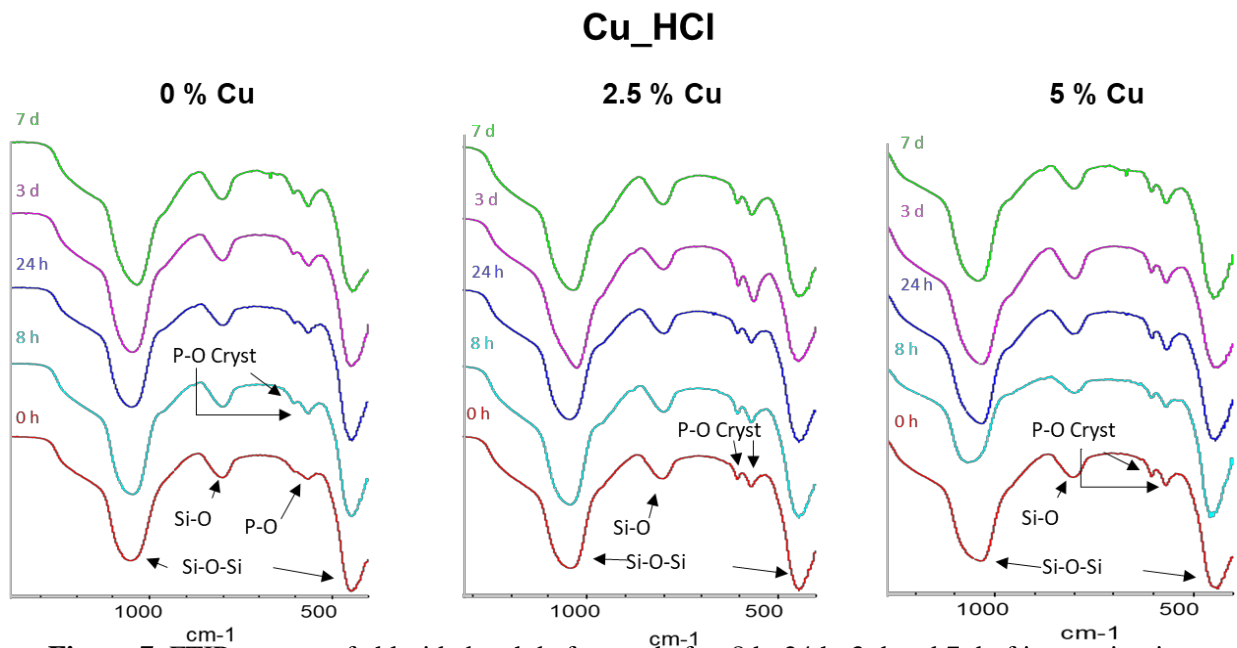

**Figure 7.** FTIR spectra of chloride batch before and after 8 h, 24 h, 3 d and 7 d of immersion in simulated body fluid. Arrows highlight the most significant bands of Si-O and P-O in the spectra. Of interest are the bands of phosphate in a crystalline environment, labeled as P-O Cryst.

(1.5-column)

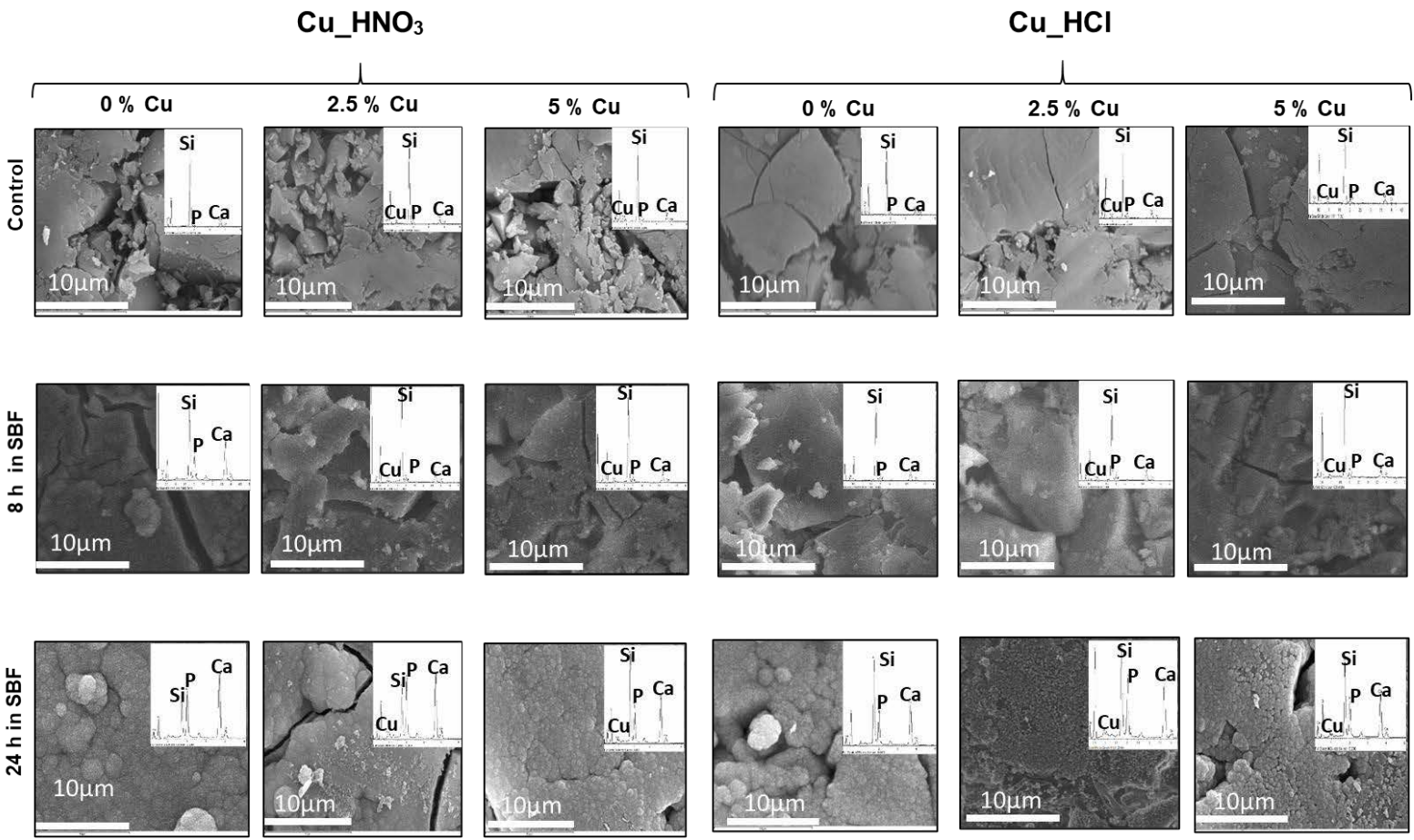

**Figure 8.** SEM images of the disks surface before and after 8 and 24 h of immersion in SBF. Each image includes their respective EDS spectrum.

(2-column)

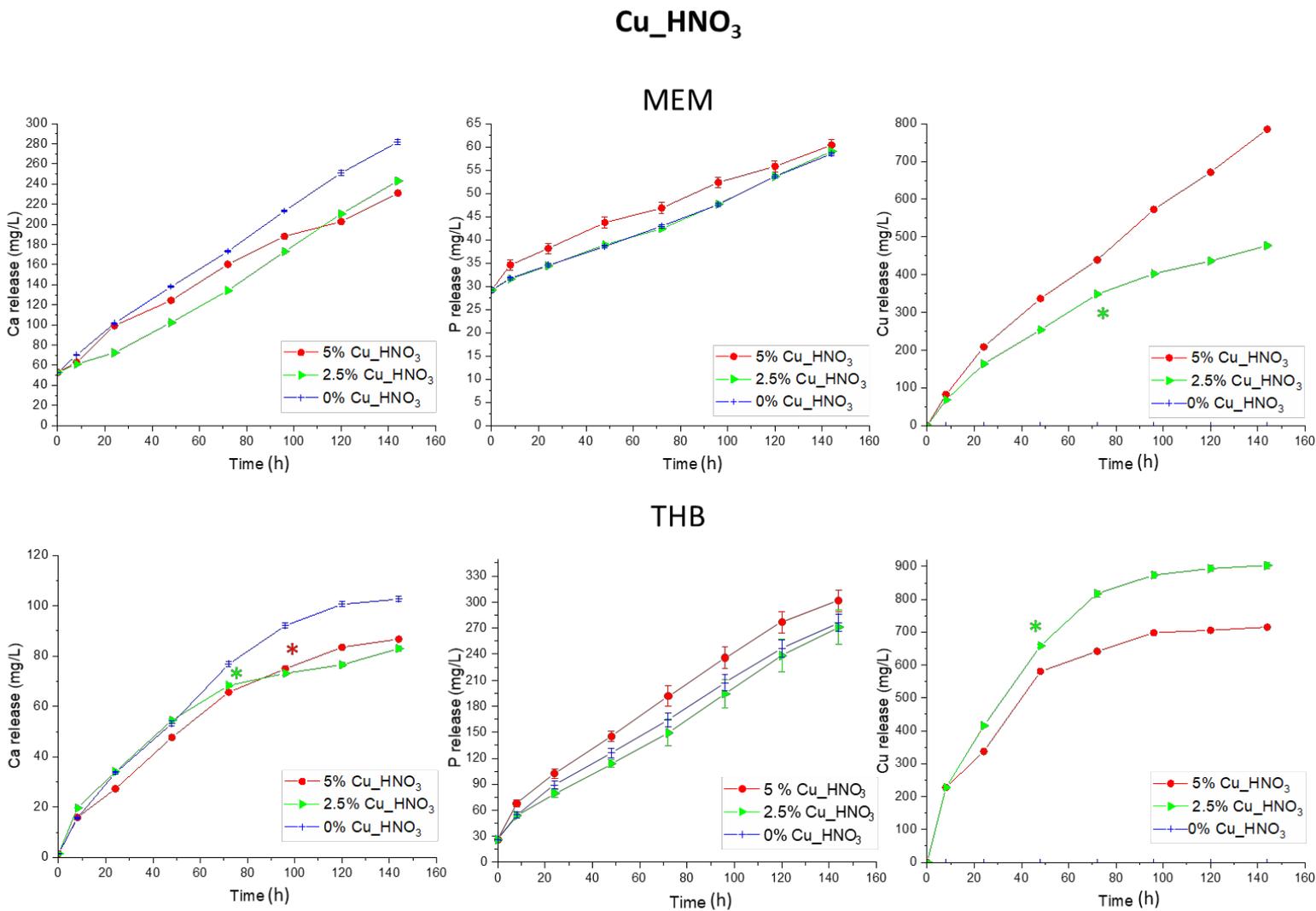

**Figure 9.** Cumulative release of calcium, phosphorous and copper from nitrate batch as a function of the time soaked in cells (MEM) and bacteria (THB) culture media. The significant differences are marked with the symbol x in the time from which a p-value>0.05 was obtained with respect to the sample without copper, except for the measurement of copper.

(2-column)

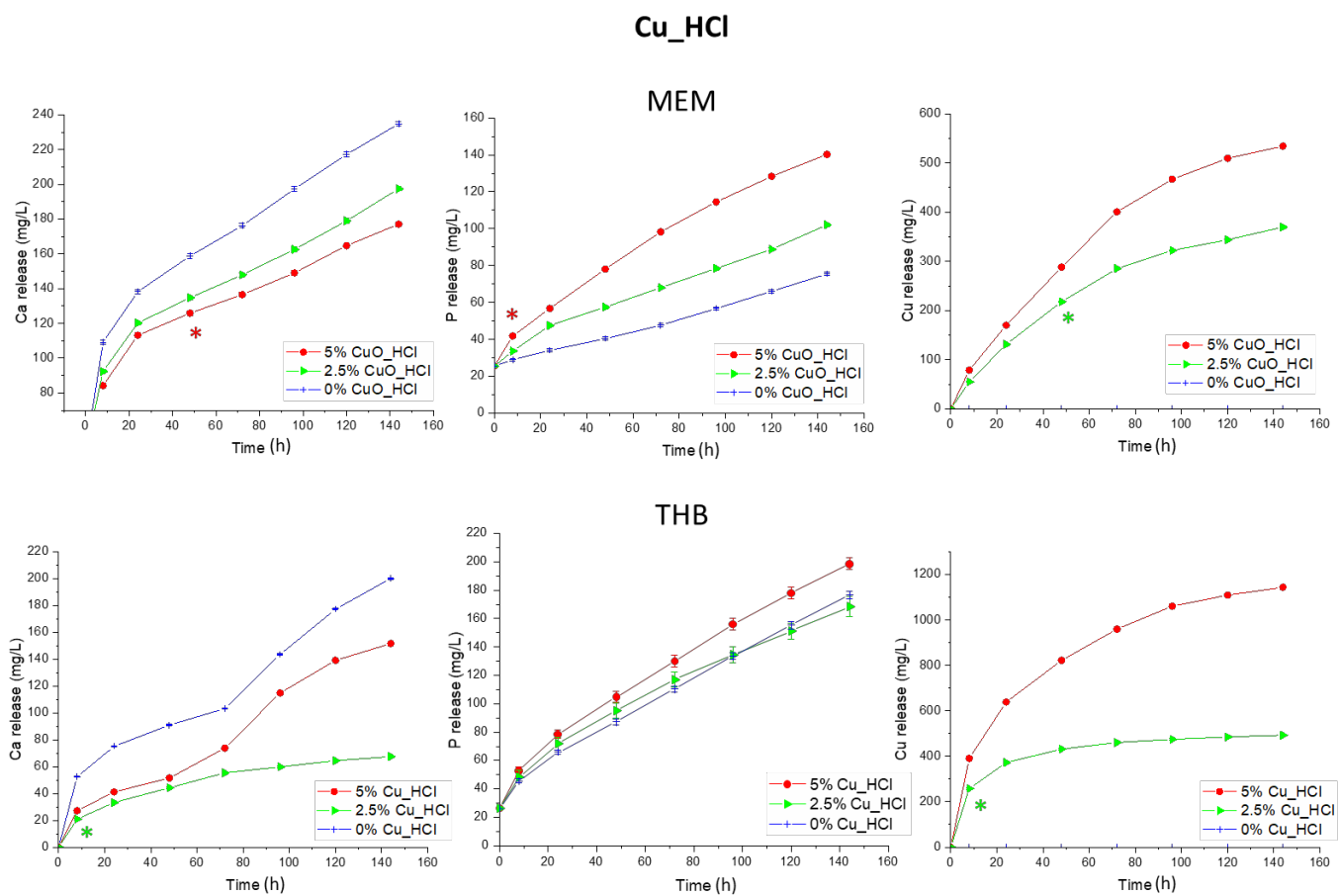

**Figure 10.** Cumulative release of calcium, phosphorous and copper from chloride batch as a function of the time soaked in cells (MEM) and bacteria (THB) culture media. The significant differences are marked with the symbol x in the time from which a p value less than 0.05 was obtained with respect to the sample without copper, except for the measurement of copper.

(2-column)